\documentclass[sigconf]{acmart}
\AtBeginDocument{%
  \providecommand\BibTeX{{%
    \normalfont B\kern-0.5em{\scshape i\kern-0.25em b}\kern-0.8em\TeX}}}

\usepackage{multirow}
\usepackage{multicol}
\usepackage{colortbl}
\usepackage{arydshln}
\usepackage{booktabs}
\usepackage{subfigure}
\usepackage{float}

\begin{document}

\title{Cross-domain User Preference Learning for Cold-start Recommendation}

\author{Huiling Zhou*, Jie Liu*, Zhikang Li*, Jin Yu, Hongxia Yang}
\thanks{*These authors contributed equally to this research.}
\email{{zhule.zhl, sanshuai.lj, zhikang.lzk, kola.yu, yang.yhx}@alibaba-inc.com}
\affiliation{%
  \institution{DAMO Academy, Alibaba Group}
 \country{China}
}

\renewcommand{\shortauthors}{Huiling Zhou, et al}

\begin{abstract}
Cross-domain cold-start recommendation is an increasingly emerging issue for recommender systems. 
Existing works mainly focus on solving either cross-domain user recommendation or cold-start content recommendation.
However, when a new domain evolves at its early stage, it has potential users similar to the source domain but with much fewer interactions. It is critical to learn a user's preference from the source domain and transfer it into the target domain, especially on the newly arriving contents with limited user feedback.
To bridge this gap, we propose a self-trained Cross-dOmain User Preference LEarning (COUPLE) framework, targeting cold-start recommendation with various \textit{semantic tags}, such as attributes of items or genres of videos.
More specifically, we consider three levels of preferences, including user history, user content and user group to provide reliable recommendation.
With user history represented by a domain-aware sequential model, a frequency encoder is applied to the underlying tags for user content preference learning.
Then, a hierarchical memory tree with orthogonal node representation is proposed to further generalize user group preference across domains. 
The whole framework updates in a contrastive way with a First-In-First-Out (FIFO) queue to obtain more distinctive representations. 
Extensive experiments on two datasets demonstrate the efficiency of COUPLE in both user and content cold-start situations. By deploying an online A/B test for a week, we show that the Click-Through-Rate (CTR) of COUPLE is superior to other baselines used on Taobao APP. Now the method is serving online for the cross-domain cold micro-video recommendation.
\end{abstract}

%
%


\settopmatter{printacmref=false}



\maketitle

\section{Introduction}
\begin{figure}[]
    \centering
    \includegraphics[width=1.0\linewidth]{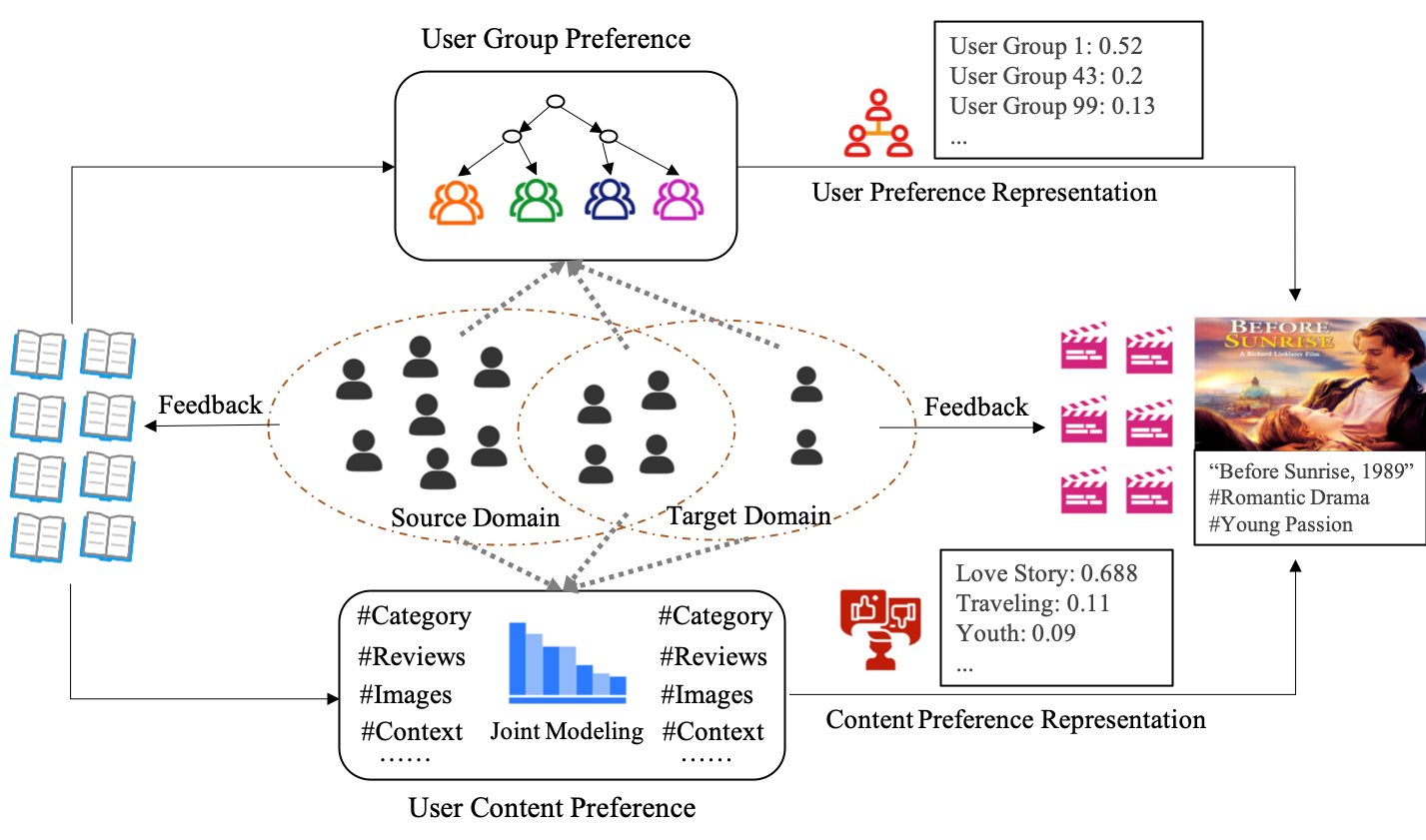}
    \caption{Overview of the proposed cross-domain cold-start recommendation system. A user's preference is captured from different perspectives with jointly modeling behaviors in all domains.}
    \label{fig:intro}
\end{figure}

Personalized recommendation system (RS) plays a vital role in an e-commerce platform, where effective strategies have been made to alleviate information overload and facilitate better user experiences. In recent years, deep learning has been widely applied in RS to overcome obstacles of conventional recommendation techniques. Great effort has been made to achieve better performances in heterogeneous multi-modal recommendation \cite{geng2015learning,epasto2019single,pal2020pinnersage}, debiased recommendation \cite{guo2020debiasing,zhou2020contrastive}, reliable and explainable recommendation  \cite{wang2019kgat,chen2020towards}, etc. 
Among all, cross-domain recommendation \cite{elkahky2015multi,hu2018conet,zhao2020catn} and cold-start recommendation \cite{li2019zero,cohen2017expediting,lu2020meta} problems draw a lot of attention. Cross-domain recommendation systems aim to transfer knowledge available in other domains (known as the source domain) to the target domain where users have much sparser feedback or interactions. Overlapped users are often selected to learn the mapping of interest between two domains so that same pattern can be applied to those cold-start users of target domain \cite{hu2018conet,zhao2020catn}. 
On the other hand, cold-start recommendation is often referred to cold-start content recommendation where no or few user feedback can be found when new items arrive. It is challenging to recommend such items to users due to the unavailability of the statistical information which RS relies heavily on. Common solutions extract and analyze content information of items to match users' potential interests based on their historical behaviors \cite{cohen2017expediting, geng2015learning}. 

Previous works normally treat cold-start user and cold-start content recommendation as two separate scenarios. However, when a new domain evolves at its early stage in real-world, it may suffer both issues at the same time. 
For example, Taobao is the largest e-commerce platform in China where billions of products are interacted with hundreds of millions of users each day. When micro-videos were first served to the platform, they shared the same set of users as product domain but had few (or none) user interactions. 
How to perform reliable recommendation on such emerging new contents appears to be one of the most challenging problems for RS. A recent work \cite{xie2020internal} proposes an internal contextual attention network to deal with the cross-domain cold-start recommendation on contents with very few interactions. However, this approach has limited  ability to model brand new contents which are never seen before.

In this paper, we further extend the work \cite{xie2020internal} and provide a unified Cross-dOmain User Preference LEarning (COUPLE) framework where items with limited interactions can be reliably recommended. 
COUPLE differs from previous cross-domain works in that we jointly model all users' behaviors across domains instead of using just the overlapped users, as shown in Figure.\ref{fig:intro}. With historical feedback modeled, a user's preference across domains is obtained from content perspective and user group perspective. When a new item comes, we push it to the corresponding users with matched interest based on its content feature and the user's general preference learnt by the system.
Main contributions of our proposed work are summarized as follows:
\begin{itemize}
\item We propose a unified user preference learning framework to solve cross-domain cold-start problem, which delicately models a user's history preference, content preference and group preference. With proper choice of a First-In-First-Out (FIFO) queue, the whole framework can be self-trained in an efficient contrastive way.
\item With semantic tags extracted for item representation, a frequency encoder based on cross-layer attention is utilized. It further boosts the interactions between high-level representation with low-level content source input. To the best of our knowledge, it is the pioneering work to solve the cross-domain cold-start issue with the use of tag information. 
\item A hierarchical memory tree with orthogonality property is proposed to learn users' preference across domains. It can generalize user's diverse interest with top $K$ leaf-node representations based on his/her historical behaviors. This is proved to be a good way of user preference generalization in cross-domain scenario.
\end{itemize}
The rest of the paper is organized as follows. In Section 2, we
review the related work. Section 3 presents the details of our proposed method. Offline experimental results and ablation study are presented in Section 4. We introduce
the online deployment of the system in Section 5 and conclude our work in Section 6.

\section{Related Work}
In this section, we introduce the related work on personalized recommendation system, user preference learning, and the orthogonal regularization mechanism we use in the paper.

\subsection{Personalized Recommendation System}
A standard matching (i.e. candidate generation) stage of a personalized recommender system aims to retrieve a small subset of items from the huge pool for the sophisticated ranking scheme. Previous methods typically employ collaborative filtering (CF) strategies to exploit user preferences from their explicit or implicit feedback \cite{hu2008collaborative,he2017neural}.
These methods achieve good performance where user-item interactions are dense but are not able to handle the cold-start situations where the interactions are limited. Content-based methods alleviate this problem by mining the content features of items or meta information about users \cite{melville2002content,geng2015learning,epasto2019single,haruna2017context,gedikli2013improving}. Among them, tags are proved to be useful \cite{gedikli2013improving,haruna2017context}, especially for cross-domain recommendation where semantically similar tags may exist in both domains and bridge the gap to some extent.
Despite their success, both types of the methods treat user-item interactions in a static way and cannot capture the dynamic interests of users over time. As a result, sequential recommender systems are proposed to model the sequential nature of users' interactions \cite{chen2018sequential,liu2016context,ma2020disentangled}. 
Besides, attention mechanism has been widely adapted to recommender systems since its out-breaking success in NLP \cite{devlin2018bert, lu2019vilbert}. It is leveraged between users and contents to capture the most representative information or provide good interpretability for recommendation \cite{kang2018self, li2019multi, cen2020controllable}.

In real-world applications, hybrid recommendation framework is usually applied which makes up for the shortcomings of using a single method and maximizes the advantages. Our work is typically within the hybrid system for cross-domain cold-start recommendation, where all strategies mentioned above are involved. 

\subsection{User Preference Learning}
The core of a recommender system is to model the dynamic user preferences. Previous works usually represent a user's historical behavior with one single latent vector which may suffer from correlation loss \cite{chen2018sequential}.  Clustering-based approaches offer an alternative to traditional model-based methods and cluster similar users or items together \cite{rafailidis2016top,liji2018improved}. However, these methods rely heavily on manual selection of a user's attribute and profile, or treat different domains separately. Recently, recommendations with memory network \cite{chen2018sequential,zhou2019topic} are proposed to model user preference and store the long term interest. The basic idea is to maintain a key-value style memory matrix with numbers of slots. A user's embedding is fed into the matrix where similarity between the user and each key vector is computed and converted to relevance probability using softmax function. In our work, a hierarchical memory tree is utilized for user preference learning. Different from previous memory-based network, our tree structure has a much larger number of memory slots (i.e. few hundred or thousand vs. few dozens). What is more, we only pick the top memory slots each time with calculated correlation score instead of using all the memory slots. 


\subsection{Orthogonal Regularization}
The orthogonality implies energy preservation and is proved to be efficient for stabilizing the distribution of activations over layers with CNNs. Orthogonal regularizers are extended to fully-connected layers and a Spectral Restricted Isometry Property (SRIP) regularizer is proposed to guarantee better convergence \cite{bansal2018can}. Later on, more investigations \cite{jia2019orthogonal,chen2019abd} further extend SRIP regularization bounds and bring compatible results on applications like image classification and Person Re-Identification. We also apply such orthogonal regularization scheme for the user group preference learning, which will be illustrated in detail in Section 3.

\section{The Proposed Method}
\begin{figure*}
    \centering
    \includegraphics[scale=0.55]{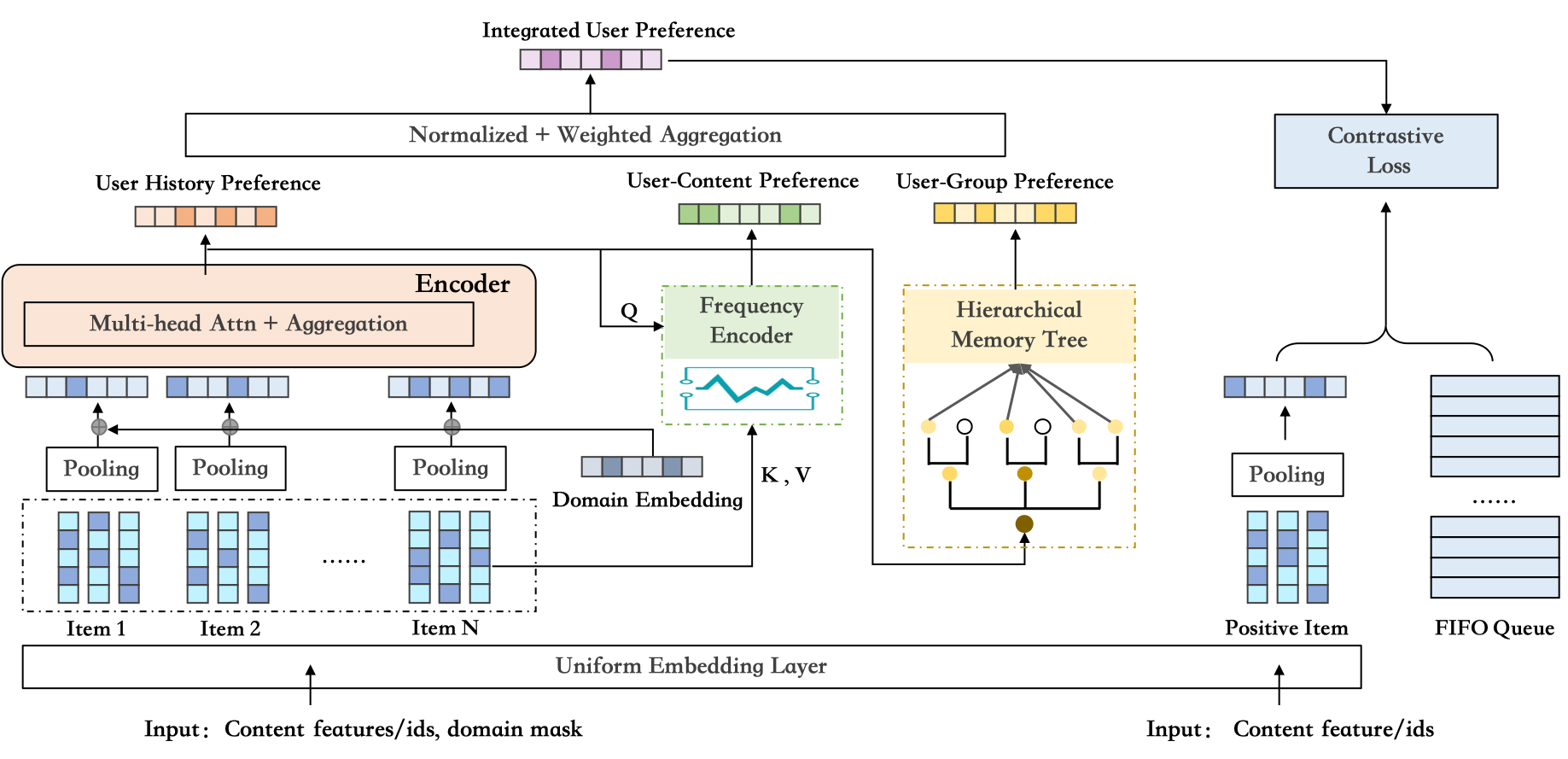}
    \caption{Framework of our proposed method. Three levels of user preferences are modeled, including user history, user content and user group. The whole framework updates in a contrastive way with a Fisrt-In-First-Out (FIFO) queue, where a large number of negative samples from latest steps is maintained for more consistant training.}
    \label{fig:framework}
\end{figure*}
In this section, we present the problem formulation of cross-domain cold-start recommendation first. Then we explain in detail how we model user preference from three different aspects in COUPLE and train the framework efficiently in a contrastive way. The overall design of our proposed COUPLE network is shown in Figure \ref{fig:framework}.

\subsection{Problem Formulation and Notations}
\subsubsection{Problem Formulation}
Cross-domain recommendation in real-world may involve several channels \cite{xie2020internal}. Without loss of generality, we classify those channels into source domain(s) \textit{\textbf{A}} and target domain(s) \textit{\textbf{B}}, where user interactions are much denser in \textit{\textbf{A}} compared to \textit{\textbf{B}}. We denote $\mathcal{U}_A$, $\mathcal{U}_B$ as the sets of users and $\mathcal{I}_A$, $\mathcal{I}_B$ as the sets of interacted items with content features (semantic tags in our case) respectively. We also have a cold-start item set in target domain denoted as $\mathcal{I}_{B}^{'}$ where items are with limited (or none) interactions. For the cross-domain cold-start recommendation, we have a) user overlap $\mathcal{U}_{A} \cap \mathcal{U}_{B} \neq \emptyset$ ; b) item overlap $\mathcal{I}_{A} \cap \mathcal{I}_{B} = \emptyset$ ; c) cold-start item overlap  $\mathcal{I}_{B} \cap \mathcal{I}_{B}^{'} = \emptyset$. We define two tasks with regard to the training and inference processes of COUPLE, respectively:
\begin{itemize}
    \item \textbf{Joint recommendation for training}, i.e., make recommendation on items in $\mathcal{I}_{A} \cup \mathcal{I}_{B}$ to users $\mathcal{U}_{A} \cup \mathcal{U}_{B}$.
    \item \textbf{Cold-start recommendation for inference}, i.e., make recommendation on items in $\mathcal{I}_{B}^{'}$ to users $\mathcal{U}_{A} \cup \mathcal{U}_{B}$.
\end{itemize}
It is noted that our work can be easily extended to recommendation on items with limited interactions as well. Notations are summarized in Table \ref{notations}.

\begin{table}[h]
  \caption{Notations Used in Section 3.}\vspace{-0.1in}
  \label{notations}
  \begin{tabular}{c|l}
    \toprule \hline
    \textbf{Notation} & \textbf{Description} \\ \hline
    $\mathcal{U}$, $\mathcal{I}$, $\mathcal{T}$ & user, item and tag set \\ \hline
    $d \in \mathbb{N}$ & latent vector dimension \\ \hline
    $n \in \mathbb{N}$ & number of tags per item \\ \hline
    $l \in \mathbb{N}$ & max sequence length \\ \hline
    $\boldsymbol{e}_u$, $\boldsymbol{e}_i$, $\boldsymbol{e}_t \in \mathbb{R}^{d \times 1}$ & user, item, tag embedding feature \\ \hline
    $\boldsymbol{e}_{u}^{(h)}, \boldsymbol{e}_{u}^{(c)}, \boldsymbol{e}_{u}^{(g)} \in \mathbb{R}^{d \times 1}$ & user history, content, group preference \\ \hline
    $\boldsymbol{s_{i, j}} \in \mathbb{R}^{d \times 1}$ & tree node vector at position $j$ in layer $i$ \\ \hline
    $\mathbf{M}$ & attention map \\ \hline
    $\mathbf{W}$ & weight matrix \\ \hline
    $\mathcal{L}$ & loss  \\ \hline
    $\|.\|$ & norm  \\ \hline
    [.] & concatenation  \\ \hline \bottomrule
  \end{tabular}
\end{table}

\subsection{Cross-domain User Preference Learning}
As illustrated in Figure \ref{fig:framework}, we model a user's preference from three interactive stages. First, user's history preference is modeled with a domain-aware multi-head attention scheme for sequential behaviors from different domains. Then it will attend to the underlying tag-level features to encode tag importance for user content preference. A hierarchical memory tree is used to represent user's historical preference with ``orthogonal" leaf-node vectors. Finally, user preferences from three aspects are weighted aggregated. 

\subsubsection{User History Preference}
We model a user's historical behavior using a deep sequential model, where item-level attention scheme is applied. In standard sequential models, a sequence of item index is often given for embedding lookup \cite{kang2018self,ma2020disentangled}. However, in cold-start content recommendation, items may be never seen or interacted before. To solve this issue, we use the global semantic tags to represent each item. To be more specific, we obtain a cross-domain tag set $\mathcal{T}$ from items across domains 
$\mathcal{I}_{A} \cup \mathcal{I}_{B}$. The tags can be extracted from hashtag made by users, item attributes or genres, or from images and texts pre-processed by vision and language models. A fair assumption can be made that the tag set $\mathcal{T}$ is able to cover most of the items in $\mathcal{I}_{B}^{'}$ from the target domain.

Our input for each item is a group of tags $[t_1,t_2,\dots,t_n]$, where $t_x \in \{1,2,\dots,N\}$ is the index of the tag in the whole set $\mathcal{T}$ with size $N$. A tag embedding table $\mathbf{H}_t \in \mathbb{R}^{d \times N}$ takes tag index sequence as the input and outputs the representation of the tag features $[\boldsymbol{e}_{t,1},\boldsymbol{e}_{t,2},\dots,\boldsymbol{e}_{t,n}]$, where $\boldsymbol{e}_{t} \in \mathbb{R}^{d \times 1}$.
In this way, we represent an item $\boldsymbol{e}_{i} \in \mathbb{R}^{d \times 1}$ by aggregating the tag features by average pooling:
\begin{equation}
\label{item_rep}
    \boldsymbol{e}_{i}=\text{AvgPool}(\{\boldsymbol{e}_{t}\}_{k=1}^{n})
    =\frac{1}{n} \sum_{k=1}^{n} \boldsymbol{e}_{t,k},
\end{equation}
where $n$ is the number of tag features capped for each item. Note that the index-based embedding lookup scheme can be easily extended to using multi-modal embeddings from any pre-trained models.
To explicitly involve the domain-specific knowledge, we follow the design in Bert \cite{devlin2018bert} and add a domain embedding table $\mathbf{H}_d \in \mathbb{R}^{d \times Q}$ ($Q$ is the number of channels) to the current item feature embedding, for training. For inference and online serving, we only use the tag features for item embedding, as suggested by the recent recommendation work \cite{zhou2020contrastive}.

To model user history preference, we apply a multi-head attention (MHAttn) mechanism \cite{kang2018self} to the sequence of temporally sorted items $[\boldsymbol{e}_{i,1},\boldsymbol{e}_{i,2},\dots,\boldsymbol{e}_{i,l}]$ clicked (rated) by users. Self-attention is applied to aggregate all item embeddings with adaptive weights and followed by two-layer feed-forward networks (FFN) to increase non-linearity.  
\begin{equation}
    (\boldsymbol{e}_{u,1},\dots,\boldsymbol{e}_{u,m})=\text{MHAttn}(\{\boldsymbol{e}_{i,k}\}_{k=1}^{l}),
\label{mhattn}
\end{equation}
where $l$ is the max sequence length and $m$ is the number of attention heads. Furthermore, we perform weighted aggregation (WA) to obtain user history preference $\boldsymbol{e}_{u}^{(h)}$ on the output representations $\{\boldsymbol{e}_{u}\}_{k=1}^{m}$ of Eq. \eqref{mhattn} as:
\begin{equation}
    \boldsymbol{e}_{u}^{(h)} = \text{WA}(\{\boldsymbol{e}_{u}\}_{k=1}^{m})
    =\text{WA}([\boldsymbol{e}_{u,1},\dots,\boldsymbol{e}_{u,m}]),
\label{wa}
\end{equation}
with a parameter matrix $\mathbf{M} \in \mathbb{R}^{d \times d}$ learnt and updated, the following procedure is performed:
\begin{equation}
\begin{split}
    &\boldsymbol{e}_m=\frac{1}{m} \sum_{k=1}^{m} \boldsymbol{e}_{u,k}, \quad
    \boldsymbol{d}_{k}=\boldsymbol{e}_{u,k}^T\mathbf{M}\boldsymbol{e}_m,\\
    &a_{k}=\frac{\exp \left(\boldsymbol{d}_{k}\right)}{\sum_{k^{\prime}=1}^{m} \exp \left(\boldsymbol{d}_{k^{\prime}}\right)},\quad
    \boldsymbol{e}_{u}^{(h)}=\sum_{k=1}^{m} a_{k} \boldsymbol{e}_{u,k}.
    \label{history_rep}
\end{split}
\end{equation}

\subsubsection{User Content Preference}
User history feature obtained from Section 3.2.1 focuses on the item-level attention, where each tag contributes equally for an item representation. However, same tags appearing multiple times in the item sequence are supposed to contribute more to a user's interest. The similar conclusion on the importance of item frequency for next-bucket recommendation has been made recently in \cite{hu2020modeling}. As a result, we propose a frequency encoder which emphasizes the appearances of the same tags. It is implemented by a cross-layer attention process.

\begin{figure}[h]
\setlength{\belowcaptionskip}{-0.2in}
    \centering
    \includegraphics[width=1.0\linewidth]{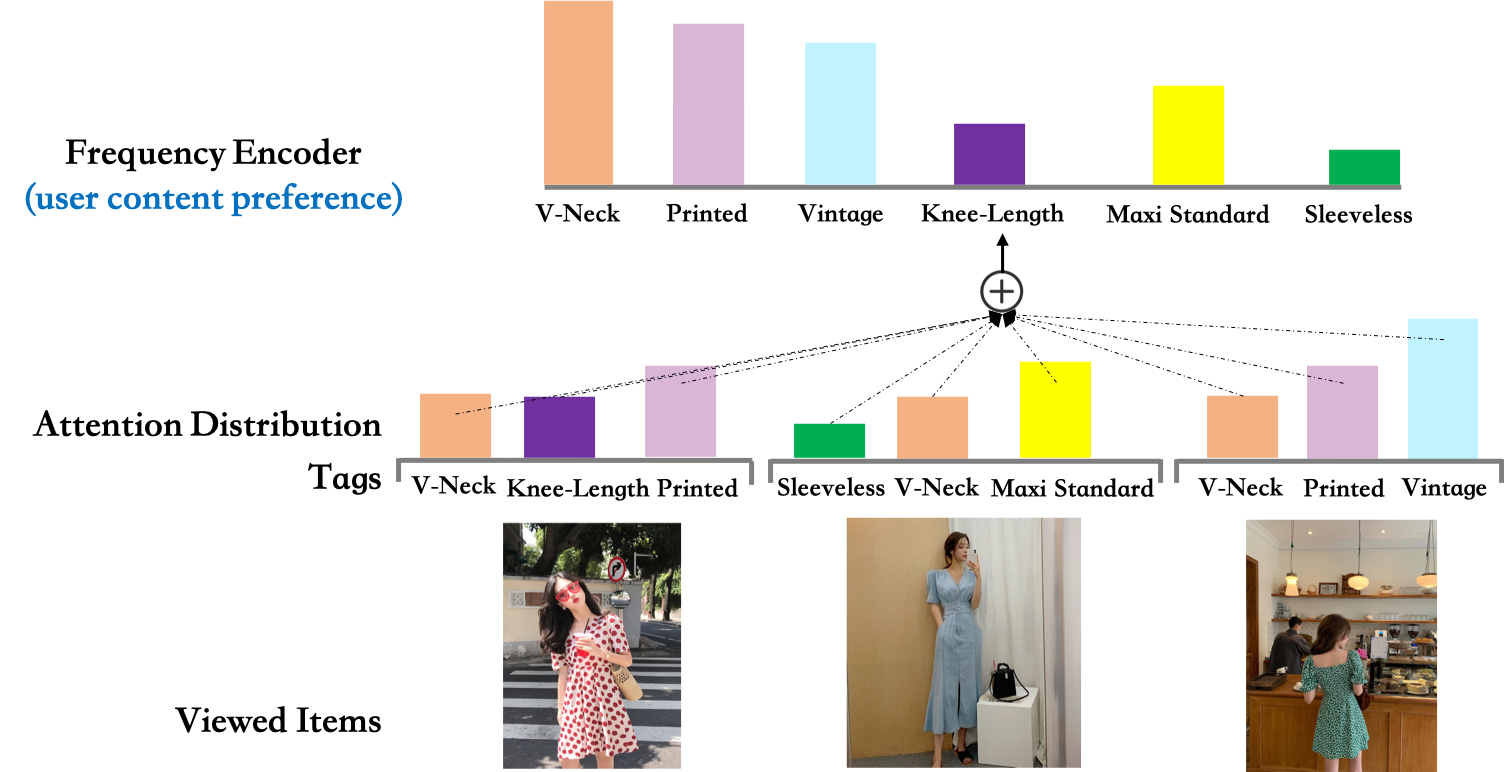}
    \caption{Tag-level Frequency Encoder. Representation of the same tag will be aggregated and augmented after user-tag attention scheme (best viewed in color).}
    \label{fig:attention}
\end{figure}

We concatenate all the tag embeddings in the item sequence together as $\{\boldsymbol{e}_{t}\}_{k=1}^{n\times{l}}$=$[\boldsymbol{e}_{t,1},\boldsymbol{e}_{t,2},\dots,\boldsymbol{e}_{t,n\times{l}}]$, and let them go through a one-layer MLP with a hyperbolic tangent function. This results in a hidden representation $\boldsymbol{h}_{t}$. Then the softmax function is applied between $\boldsymbol{h}_{t}$ and the user's history representation $\boldsymbol{e}_{u}^{(h)}$ to obtain the attention weights. The user content representation $\boldsymbol{e}_{u}^{(c)}$ is computed as the weighted sum of attention weights and the tag features:
\begin{equation}
\begin{split}
    &\boldsymbol{h}_{t,i}=\operatorname{tanh}(\textbf{W}_{w}\boldsymbol{e}_{t,i}+b_w),\quad
    a_{i}=\frac{\exp \left(\boldsymbol{e}_{u}^{(h)} \cdot \boldsymbol{h}_{t,i} \right)}{\sum_{i^{\prime}=1}^{n\times{l}} \exp \left(\boldsymbol{e}_{u}^{(h)} \cdot \boldsymbol{h}_{t,i^{\prime}}\right)}, \\
    &\boldsymbol{e}_{u}^{(c)}=\sum_{i=1}^{n\times{l}} a_{i} \boldsymbol{e}_{t,i}.
\end{split}
\label{content}
\end{equation}
In this way, the same tag with multiple attention weights will be augmented in the final representation of user content preference. A simple illustration of how this frequency encoder works is shown in Figure.\ref{fig:attention}. From the design point of view, the user history representation is modeled based on a bottom-up attention while the user content preference is in a top-down fashion. Such cross-layer attention is proved to be critical to boost interactions between the high-level representation with the low-level content source  \cite{anderson2018bottom,lu2019vilbert}. 

\subsubsection{User Group Preference}
 Unlike previous cross-domain works where only the overlapped users are considered, we take into account all the users’ behavior patterns and try to find out a uniform representation for user group preference across domains. Consequently, a hierarchical memory tree is designed to remember a user's general interest and automatically classify it into groups with closest interests. It differs from the original memory networks in that : a) we are able to have more memory slots as proved in the recent work \cite{shi2020beyond} (i.e. few hundreds or thousands vs. few dozens); and b) we don't use all slots as memory networks do, instead we pick the top $K$ leaf-node slots each time with highest scores. 

\textbf{Tree Structure Design.} With user history representation from equation \eqref{history_rep}, we follow the similar procedure as \cite{shi2020beyond} to use a hierarchically fully-connected tree where the root node allocates weights to children nodes and finally leaf nodes. To enforce further interpretability, the sum of the node weights at the same layer is equal to 1 and the sum of children node weights inherited from the same parent node is equal to the parent node weight. Given user history representation $\boldsymbol{e}_{u}^{(h)}$ and a parent node at position $j$ in layer $i$ with $h$ children nodes, whose weight is denoted as ${w}_{i,j}$, the weights of the children nodes in layer $i+1$ can be calculated using softmax function:
\begin{equation}
\begin{split}
    w_{i+1,k}
    &={w}_{i,j} \cdot \operatorname{softmax}(\boldsymbol{e}_{u}^{(h)} \cdot \boldsymbol{s}_{i+1,k}) \\
    &={w}_{i,j} \cdot \frac{\exp (\boldsymbol{e}_{u}^{(h)} \cdot \boldsymbol{s}_{i+1,k})}{\sum_{l} \exp (\boldsymbol{e}_{u}^{(h)} \cdot \boldsymbol{s}_{i+1,l}))}, \quad \forall l=1,2,\cdots,h
\end{split}
\label{nodeweight}
\end{equation}
where $\boldsymbol{s}_{i+1,k}$ is the memory slot vector at position $k$ of layer $i+1$.

\textbf{Dead-node Solution.} During the training process of retrieving top $K$ leaf-node vectors, we met the same problem as \cite{shi2020beyond} where almost the same set of nodes where visited at each step. The paper dealt with this problem by uniformly choosing $K$ random candidates each time thus solving the dead-node problem at the early stage of training. However, as the model parameters update along time, real top-$K$ node vectors are desired to be retrieved.

In our work, we use Gumbel-Softmax \cite{jang2016categorical} which is an ideal sampling trick for distribution on discrete vectors (also known as reparameterization trick). The basic idea of Gumbel-Softmax is to add a Gumbel noise and temperature factor to the original softmax function so that it is able to approximate a probability distribution made up of discrete categories (which are the normalized leaf-node weights in our case).

With calculated leaf-node weights (i.e. probabilities) $w_1,w_2,\dots,w_k$ from softmax function using Eq. \eqref{nodeweight}, the Gumbel-Softmax is conducted as:
\begin{equation}
\begin{split}
    y_{i}
    &=\operatorname{softmax}((\operatorname{log}({w_i})+g_i)/\tau) \\
    &=\frac{\exp ((\operatorname{log}({w_i})+g_i)/\tau)}{\sum_{j} \exp ((\operatorname{log}({w_j})+g_j)/\tau)}, \quad \forall j=1,2,\cdots,k
\end{split}
\label{gumbel}
\end{equation}
where $g_1,g_2,\dots,g_k$ are i.i.d samples drawn from the standard Gumbel distribution having $\mu$ and $\beta$ as 0 and 1 respectively, with PDF (Probability Density Function) of $e^{-(x+e^{-x})}$. In practice, $g_k$ can be sampled using inverse transform sampling by drawing $u_k\sim \text{uniform}(0,1)$ and calculated as $g_k=-\log(-\log(u_k))$ \cite{maddison2016concrete}.

One of the great property of the Gumbel-Softmax is that the output value approaches the real distribution with low temperature factor $\tau$ and tends to be uniform sampling with larger $\tau$. As a result, we start the training with a big temperature and then anneal it towards small values. In this way, the model tends to explore random nodes for update in the beginning and gradually stick to the real top $K$ selection for a better convergence.

For inference, the Top $K$ leaf-node vectors with highest weight scores are used to form the user group representation $\boldsymbol{e}_{u}^{(g)}$:
\begin{equation}
    \boldsymbol{e}_{u}^{(g)}=\sum_{k=1}^{K} w_{k}^{(leaf)} \boldsymbol{s}_{k}^{(leaf)}.
\end{equation}

\textbf{Orthogonality}. To make the nodes at each layer more diverse and representative, orthogonality regularizer is applied to the fully-connected layers $\mathbf{W}$ in the tree:
\begin{equation}
\label{orthogonal}
    \mathcal{L}_O=\lambda \cdot \sigma(\mathbf{W}^{T} \mathbf{W}-\mathbf{I}),
\end{equation}
where $\lambda$ is the penalty parameter and $\sigma(\mathbf{W})=\sup _{x \in \mathbb{R}^{1 \times n}, x \neq 0} \frac{\|\mathbf{W} x\|}{\|x\|}$ is the spectral norm of $\mathbf{W}$, i.e. the largest singular value of $\mathbf{W}$. Though computation of Eq. \eqref{orthogonal} involves expensive eigen-decomposition, it can be approximated via power iteration method \cite{bansal2018can}. Starting with a random initialized vector $\boldsymbol{v} \in \mathbb{R}^{d}$, we 
iteratively perform the following procedures a small number of times (2 by default):

\begin{equation}
    \boldsymbol{u} \leftarrow(\mathbf{W}^{T} \mathbf{W}-\mathbf{I}) \boldsymbol{v}, \boldsymbol{v} \leftarrow(\mathbf{W}^{T} \boldsymbol{W}-\mathbf{I}) \boldsymbol{u}, \sigma(\mathbf{W}^{T} \mathbf{W}-\mathbf{I}) \leftarrow \frac{\|\boldsymbol{v}\|}{\|\boldsymbol{u}\|}.
\end{equation}

Our hierarchical tree-based memory module is similar to previous work \cite{shi2020beyond} in design. However, the main purpose and realization is quite different:
\begin{itemize}
\item The PreHash module proposed in \cite{shi2020beyond} is designed to solve user embedding problem where certain anchor vectors are manually selected and kept unchanged during training. In our work, the memory tree is designed for user group classification and is fully automatic with parameters updating. All the user behavior representations will go through this module and fall into most related leaf slots.
\item Orthogonality property is applied to our memory tree as it is a strong regularization for the diverse representation and ensures better convergence of our tree module while PreHash in \cite{shi2020beyond} doesn't have this for model update.
\item Sampling strategy during training is also different. Prehash uses the uniform sampling throughout training while the Gumbel-softmax sampling applied in COUPLE is more efficient for node representation update.
\end{itemize}
After obtaining user history representation (in Section 3.2.1), user content representation (in Section 3.2.2) and user group representation (in Section 3.2.3), we apply weighted aggregation (WA) again to get the final user representation $\boldsymbol{e}_{u}$ using Eq. \eqref{wa}: 
\begin{equation}
    \boldsymbol{e}_{u}=\text{WA}([e_{u}^{(h)},e_{u}^{(c)},e_{u}^{(g)}]).
\end{equation}

\subsection{Contrastive Learning}
Unsupervised representation learning in a contrastive way is highly successful in recently research \cite{radford2018improving,devlin2018bert,he2020momentum}. Following a standard contrastive definition \cite{hadsell2006dimensionality}, we define a contrastive loss between our user representaion $\boldsymbol{e}_{u}$ and item representation $\boldsymbol{e}_{i,0}$,$\boldsymbol{e}_{i,1}$,...,$\boldsymbol{e}_{i,m}$:
\begin{equation}
    \mathcal{L}_{N}=-\mathbb{E}[\log \frac{\exp (\boldsymbol{e}_{u}\cdot\boldsymbol{e}_{i}^{+}/\omega)}{\exp (\boldsymbol{e}_{u}\cdot \boldsymbol{e}_{i}^{+}/\omega)+\sum_{j=1}^{m-1} \exp (\boldsymbol{e}_{u}\cdot \boldsymbol{e}_{j}^{-}/\omega)}],
\label{contrastive}
\end{equation}
where $\omega$ is a temperature hyper-parameter. The contrasitve loss has a low value when $\boldsymbol{e}_{u}$ is similar to its positive item sample $\boldsymbol{e}_{i}^{+}$ and dissimilar to all other negative item samples $\{\boldsymbol{e}_{j}^{-}\}$.

In earlier study, softmax-based classifier is often utilized to classify between positive and negative samples. In \cite{he2020momentum}, an efficient momentum contrastive scheme is proposed to maintain the dictionary as a queue of data samples. In this way, a rich set of negative samples can be involved for training and achieve positive results in various computer vision tasks. In our framework, a First-In-First-Out (FIFO) quque is maintained and updated with the current batch enqueued and the oldest batch dequeued for each training step. So that each batch of positive samples can be encoded with the hard negative samples over the latest steps. It makes the loss tracking more consistent and has a debiasing effect in large-scale production environment as proved in \cite{zhou2020contrastive}. Thus, our network is trained under the loss function $\mathcal{L}=\mathcal{L}_N+\mathcal{L}_O$ consisting of a contrastive loss and a orthogonal constraint on weights of memory tree.

\section{Experiments}
In this section, we evaluate our proposed cross-domain user preference learning (COUPLE) model for top-$K$ retrieval task under different scenarios. We introduce the experimental setup and the comparing baselines first, and present the performance comparison with other state-of-the-arts. Then the ablation study is carried out both quantitatively and qualitatively.

\subsection{Experimental Setup}
\subsubsection{Dataset Overview}
We compare the performance of our proposed framework with related methods for cross-domain cold-start recommendation on two publicly accessible datasets. Amazon review dataset\footnote{http://jmcauley.ucsd.edu/data/amazon/} is one of the most widely-used public dataset \cite{he2016ups} for e-commerce recommendations. We use subsets of Books as the source domain while Movies and TV the target domain.
The other dataset is called Tao-Product-Micro-Video (TPMV), selected from the real-world online service of Taobao Recommendation. User interactions are much denser in product domain than micro-video domain, making the dataset suitable for cross-domain cold-start recommendation. The detailed statistics of the two datasets are demonstrated in Table \ref{dataset}.

\begin{table}[]
\renewcommand{\arraystretch}{1.2}
\centering
\caption{Dataset statistics.}
\label{dataset}
\begin{tabular}{c|c|c|c|c}
\toprule 
Datasets & \#Users & \#Item & \#Tags & \#Interact\\
\midrule
Amazon Book& 1,672,201 & 1,677,035 & 555,685 & 11,637,960\\ \hline
Amazon Movie& 1,080,338 & 154,451 & 73,430 & 2,561,003\\ \hline
Taobao Product& 951,069 & 2,794,416 & 85,398 & 158,758,876\\ \hline
Taobao m-Video& 829,930 & 969,145 & 56,291 & 47,332,179\\ \bottomrule
\end{tabular}
\vspace{-1.5mm}
\end{table}

\subsubsection{Experimental Setup}
To compose valid training and testing datasets for cross-domain cold-start recommendation, we use "click" as the implicit feedback for TPMV dataset and "review rating" for Amazon dataset. For training and evaluation, clicked items of TPMV and review ratings over 3 (ratings range from 1 to 5) of Amazon are regarded as positive samples and temporally sorted with timestamps of a user's action. For testing, we use leave-one-out strategy and make the last-position items from target domain as the ground-truth, while erasing these items from the training pool to make sure they are totally cold-start contents. To further make the situation tougher as the real circumstances, we randomly select half of the testing users, and erase all the items 
of the target domain (if any) from their historical sequences to make them cold-start users. For all the models compared in this paper, the detailed numbers of training and testing samples are listed in Table \ref{traintest}.

\begin{table}[h]
\renewcommand{\arraystretch}{1.2}
\centering
\caption{Traing and testing splits.}\vspace{-0.1in}
\label{traintest}
\begin{tabular}{c|c|c|c}
\toprule 
Datasets & \#Train & \#Test & \#Cold-start User\\
\midrule
Amazon& 5,797,557 & 95,884 & 67,063 \\ \hline
TPMV& 5,333,879 & 62,230 & 34,934\\ \bottomrule
\end{tabular}
\vspace{-2.5mm}
\end{table}\vspace{-0.1in}

\subsubsection{Evaluation metrics}
To evaluate the performance of all the methods, Hit Ratio (HR), and Normalized Discounted
cumulative gain (NDCG) are used. HR measures whether the positive item retrieved within the top-$K$ and NDCG penalizes the score if positive item positioned lower in the ranking list. Higher values in these metrics indicate better recommendation performance.
For each ground-truth positive item, we pair it with 100 randomly sampled negative items from the pool where users have no historical interactions with, following the prior work \cite{he2017neural,huang2018improving}. To make the comparison more reliable, 50 of them are sampled randomly, while the other 50 items are sampled according to the popularity \cite{huang2018improving}.

\begin{small}
\begin{table*}[t]
  \centering
  \caption{HitRate and NDCG of different methods on the two datasets, where best performance is in boldface.
HP denotes hyperparameters, including $n$ the number of tags and $l$ the item length for sequential modeling. }
  \label{hr&ndcg}
  \renewcommand{\arraystretch}{1.3}
  \begin{tabular}{c|cccc|cccc}
    \toprule
    Dataset & \multicolumn{4}{c|}{TPMV Data} & \multicolumn{4}{c}{Amazon Data} \\ 
    \midrule
    HP & \multicolumn{4}{c|}{$n$ = 15, $l$ = 5} & \multicolumn{4}{c}{$n$ = 10, $l$ = 8} \\
    \midrule
    Metric & HitRate@5 & HitRate@10 & NDCG@5 & NDCG@10  & HitRate@5 & HitRate@10 & NDCG@5 & NDCG@10 \\
    \midrule
    DSSM \cite{huang2013learning} & 0.07764 & 0.12490 & 0.05397 & 0.06908 & 0.05119 & 0.09463 & 0.03093 & 0.04658 \\ 
    DeepCoNN \cite{zheng2017joint} & 0.19080 & 0.27351 & 0.13435 & 0.16062 & 0.05352 & 0.10278 & 0.03319 & 0.04800 \\
    \cdashline{1-9}[0.8pt/2pt]
    Youtube \cite{covington2016deep} & 0.19705 & 0.28187 & 0.13899 & 0.16627 & 0.06777 & 0.12398 & 0.04174 & 0.05970 \\ 
    ICAN \cite{xie2020internal}& 0.20610 & 0.28502 & 0.14711 & 0.17248 & 0.07431 & 0.13071 & 0.04601 & 0.06405\\
    \cdashline{1-9}[0.8pt/2pt]
    SASRec \cite{kang2018self}& 0.20145 & 0.28491 & 0.14709 & 0.17154 & 0.06918 & 0.12410 & 0.04352 & 0.06109 \\
    ComiRec \cite{cen2020controllable}& 0.20171 & 0.28539 & 0.14418 & 0.17110 & 0.07036 & 0.12684 & 0.04322 & 0.06127 \\ 
    Ours & \textbf{0.22444} & \textbf{0.30625} & \textbf{0.16433} & \textbf{0.19062} & \textbf{0.07753} & \textbf{0.13358} & \textbf{0.04920} & \textbf{0.06711} \\  
    \cdashline{1-9}[0.8pt/2pt]
    Improvement & + 8.89\%  & + 7.44\% & + 11.71\% & + 10.52\% & + 4.30\% & + 2.19\% & + 6.93\% & + 4.77\% \\ 
    \bottomrule
  \end{tabular}
\end{table*}
\end{small}

\subsection{Competitors}
We compare our proposed framework with six baselines with neural deep learning structures, including DSSMs, Youtube DNNs and the Sequential methods.
\begin{itemize}
    \item \textbf{DSSMs}. The original Deep Semantic Similarity Model (DSSM) \cite{huang2013learning} serves as a strong baseline widely used in information retrieval, which can perform semantic matching between a query and a document. DeepCoNN \cite{zheng2017joint} further extends DSSM to two parallel convolutional neural networks. One of them models user behaviors and the other models item properties from the review texts. It achieves a strong performance in content recommendation.
    \item \textbf{Youtube DNNs}. Youtube \cite{covington2016deep} is a classical deep-based matching model for building user and item embeddings collaboratively. Layers of depth on the top is proved effective to model non-linear interactions between features. The latest work \cite{xie2020internal} solves multi-channel cold-start issue based on Youtube DNN structure where SOTA result is achieved.
    \item \textbf{Sequential methods}. SASRec \cite{kang2018self} encodes user's behavior based on a variant of Transformer and is used as the backbone for many sequential models including ours. And we also compared to the latest sequential method ComiRec \cite{cen2020controllable} which further improves for controllable multi-interest matching based on the previous MIND \cite{li2019multi}.
\end{itemize}
For methods mentioned above, we follow their implementations with tuned parameters for better comparison. Original DSSM \cite{huang2013learning} is implemented by maximizing the conditional likelihood of clicked items given only a closest query (item in our case). For DeepCoNN \cite{zheng2017joint}, it models similarity between user and item representations. We feed in a sequence of items represented by tag features where convolution is applied. In terms of sequential methods, we first average tag representations for an item as our method does. Then item sequence is modeled the same way as the original papers, where position embedding is added. To implement ICAN \cite{xie2020internal}, we adjust the ID input feature with tag embeddings to fit in the content cold-start scenario, while keeping all other settings unchanged.

\subsection{Performance Comparison}
The overall performance of all the methods on both datasets is summarized in Table \ref{hr&ndcg}. We can observe that our proposed COUPLE network, which models user preference from different perspectives, consistently yields the best performance in terms of HitRate@N and NDCG@N (N = 5, 10).
We would like to note that, among three different types of baselines, Youtube DNNs and Sequential methods obtain better performance than DSSMs. It implies that sufficient interactions between user and item features are crucial for content-based recommendation. Within two-towel models, DeepCoNN gains apparent improvement over original DSSM by modeling series of items for user presentation instead of only one item.
By adding channel-wise attention between source and target domain, ICAN is able to achieve better performance over original Youtube network. The same improvement can be found on ComiRec over simple sequentially self-attention method SASRec, where multi interests of users are learnt across domains.

Furthermore, we observe that the improvement of our method over the strongest baselines is more significant on TPMV data (7.44\% vs. 2.19\% on HitRate@10) and the performances of all the methods on TPMV Dataset are generally better than those on Amazon Dataset.
The reasons may be that: a) TPMV data have more overlapped tags in target domain with source domain than Amazon (60\% vs. 25\%), so that models can learn a better correlation between domains; and b) user behaviors are more consistent on TPMV data which is collected from a range of a few days compared to Amazon with a wider range of months (over 4 years). Both attributes of TPMV favors our model where user-content attention and user-group preference learning are performed to generalize and transfer the interests across domains. 

\subsection{Ablation Study}
We perform ablation study to further explore the effect of individual components of our COUPLE model on performance. As shown in Table \ref{ablation}, our base model with user history and domain embeddings achieves similar performance with SASRec. An apparent improvement is observed by applying user group preference and user content preference learning modules. The variant with hierarchical memory tree outstands others and raise the HitRate@10 by a large margin of 3.88\%, further illustrating the effectiveness of the generalizing user preference across domains. After combining all modules together to represent the user, the performance is further improved. 
We also experiment with the batch negative sampling strategy compared to the contrastive learning using a FIFO queue. It can be seen that with a larger number of negative samples, model is able to distinguish data with higher efficiency. With all modules put together, our proposed model achieve a competitive performance for the real-world cold-start recommendation on Taobao platform.

\begin{table}[h]
    \centering
    \setlength{\belowcaptionskip}{-0.1in}
    \renewcommand{\arraystretch}{1.2}
    \caption{Ablation Study on TPMV Dataset.}
    \label{ablation}
    \begin{tabular}{ccc}
        \toprule
        & \multicolumn{2}{c}{\textbf{Metrics}} \\
        \cmidrule(r){2-3}
        \textbf{Variants of Our Method} & HR@10 & NDCG@10 \\
        \midrule
        Base (User History + Domain) & 0.28411 & 0.17042 \\
        Base + User-Content Preference & 0.28928 & 0.17531 \\
        Base + Memory Tree & 0.29554 & 0.18392 \\
        All User + Batch\_neg & 0.3011 & 0.18787 \\
        \textbf{All User + FIFO Queue} & \textbf{0.30625} & \textbf{0.19062} \\
        \bottomrule
    \end{tabular}
\end{table}\vspace{-0.1in}

\subsection{Source-domain Data Sensitivity}
One critical assumption of our work is that it helps with the cross-domain recommendation by joint modeling user behaviors from both source and target domains. In this section, We conduct experiments on two datasets the sensitivity of our model against the amount of source domain data. By manipulating the amount of training data, we gradually decrease the portion of data from the source domain. As part of the testing cases is cold-user recommendation where user behaviors only exist in the source domain, we further disassemble HR@10 and NDCG@10 for different situations.

\begin{figure}[]
\setlength{\belowcaptionskip}{-0.15in}
    \centering
    \subfigure[Impact on HR of TPMV]{\includegraphics[scale=0.25]{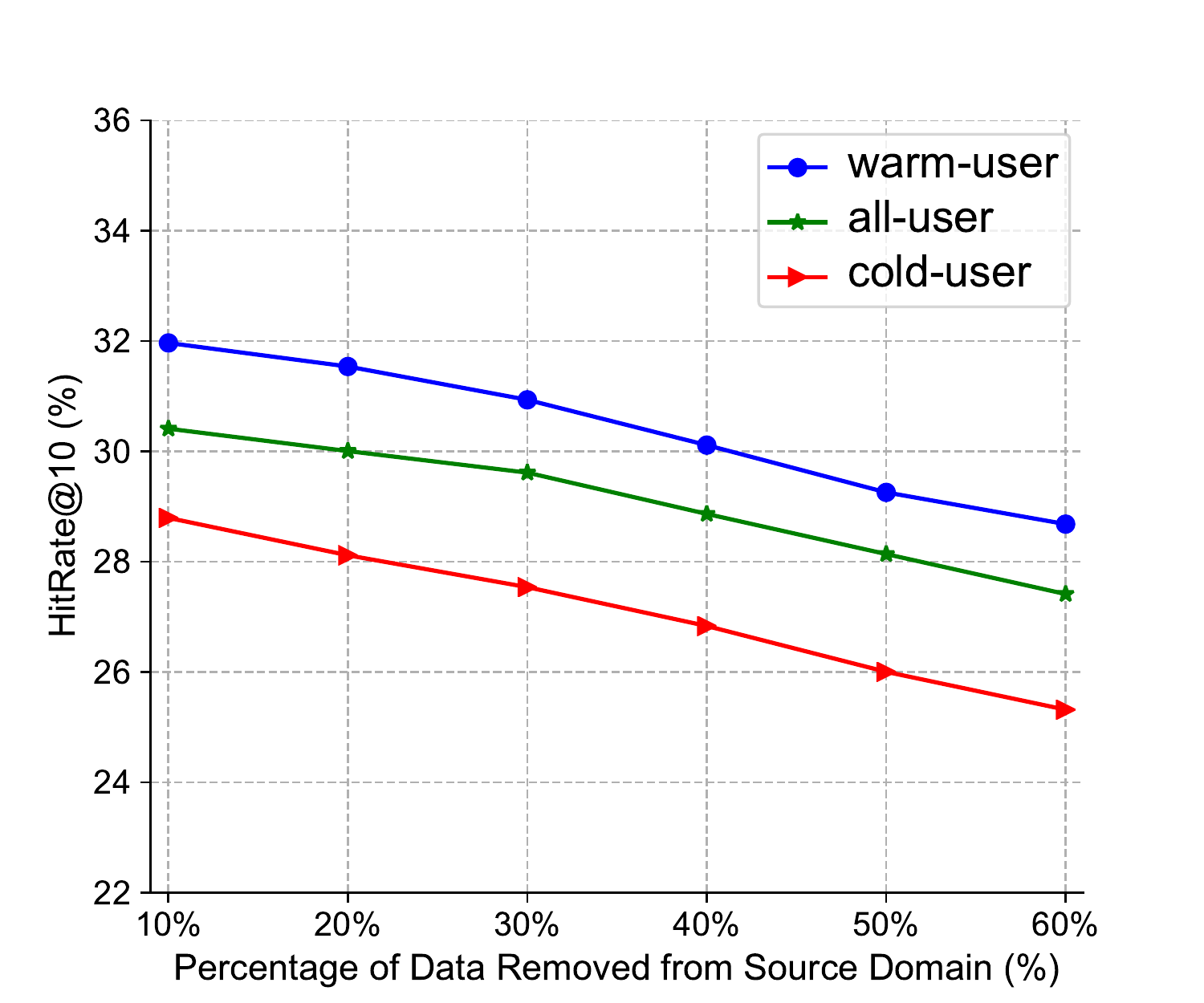}}
    \vspace{-0.15in}
    \subfigure[Impact on NDCG of TPMV]{\includegraphics[scale=0.25]{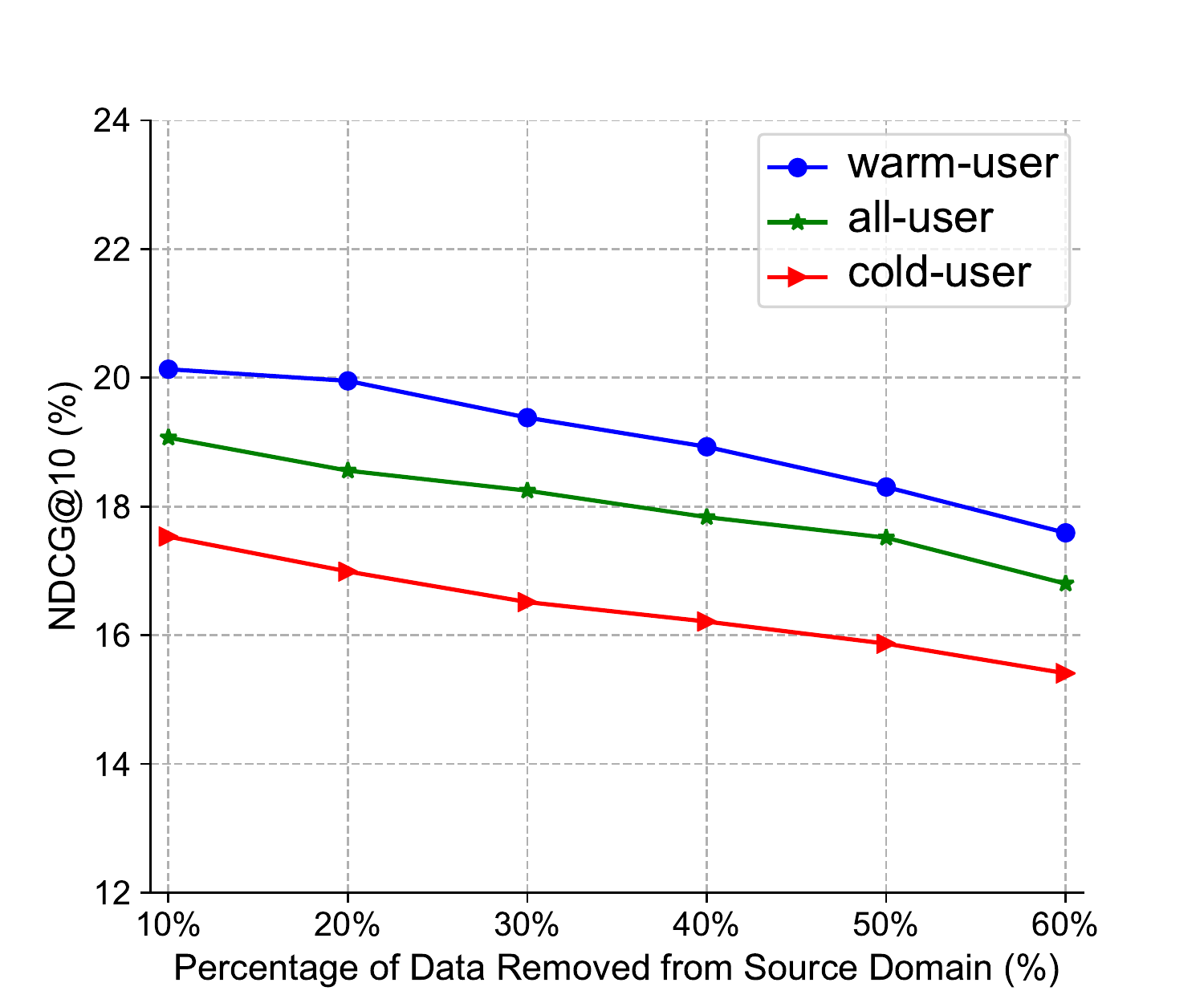}}
    \vspace{-0.15in}
    \subfigure[Impact on HR of Amazon]{\includegraphics[scale=0.25]{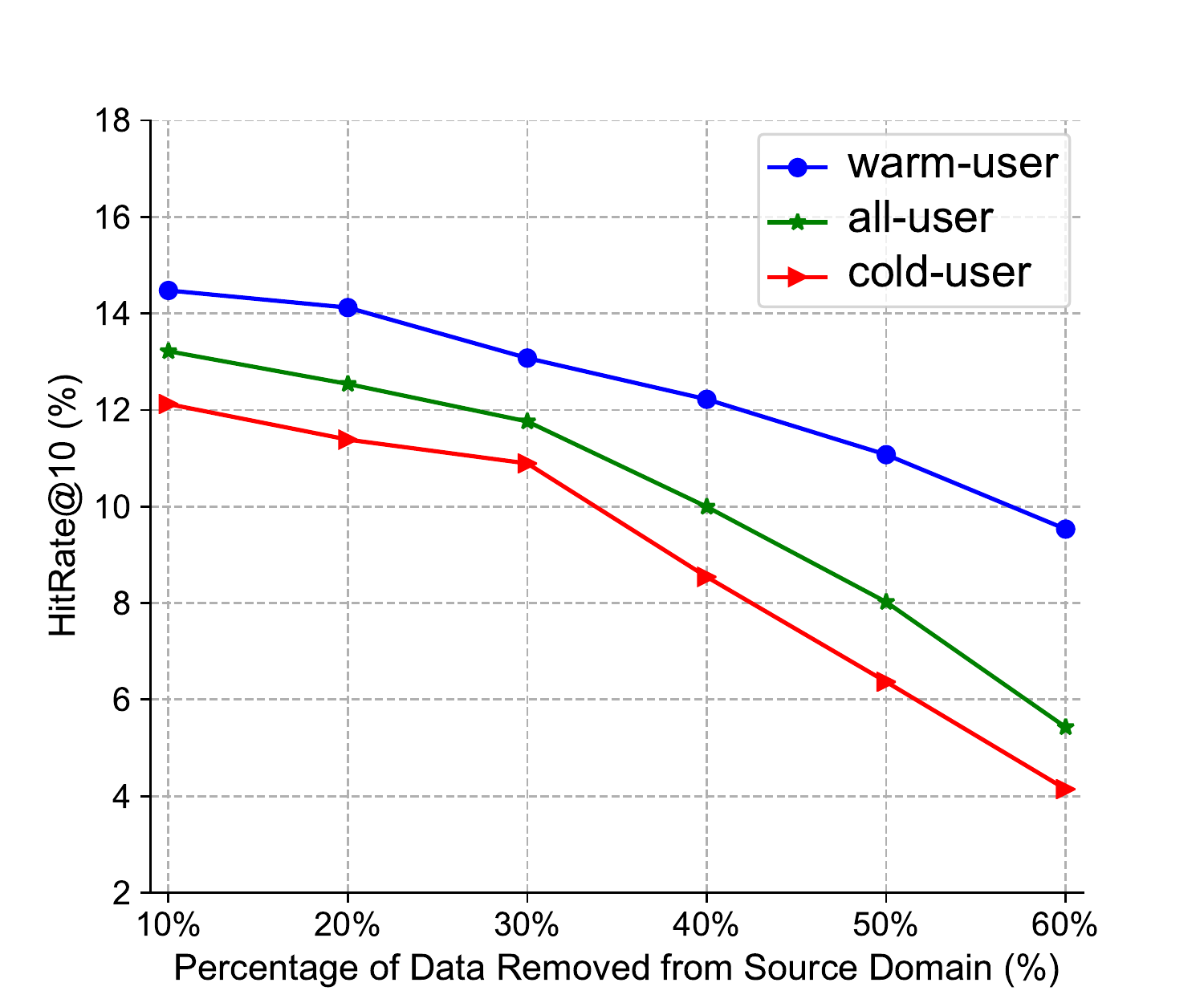}}
    \subfigure[Impact on NDCG of Amazon]{\includegraphics[scale=0.25]{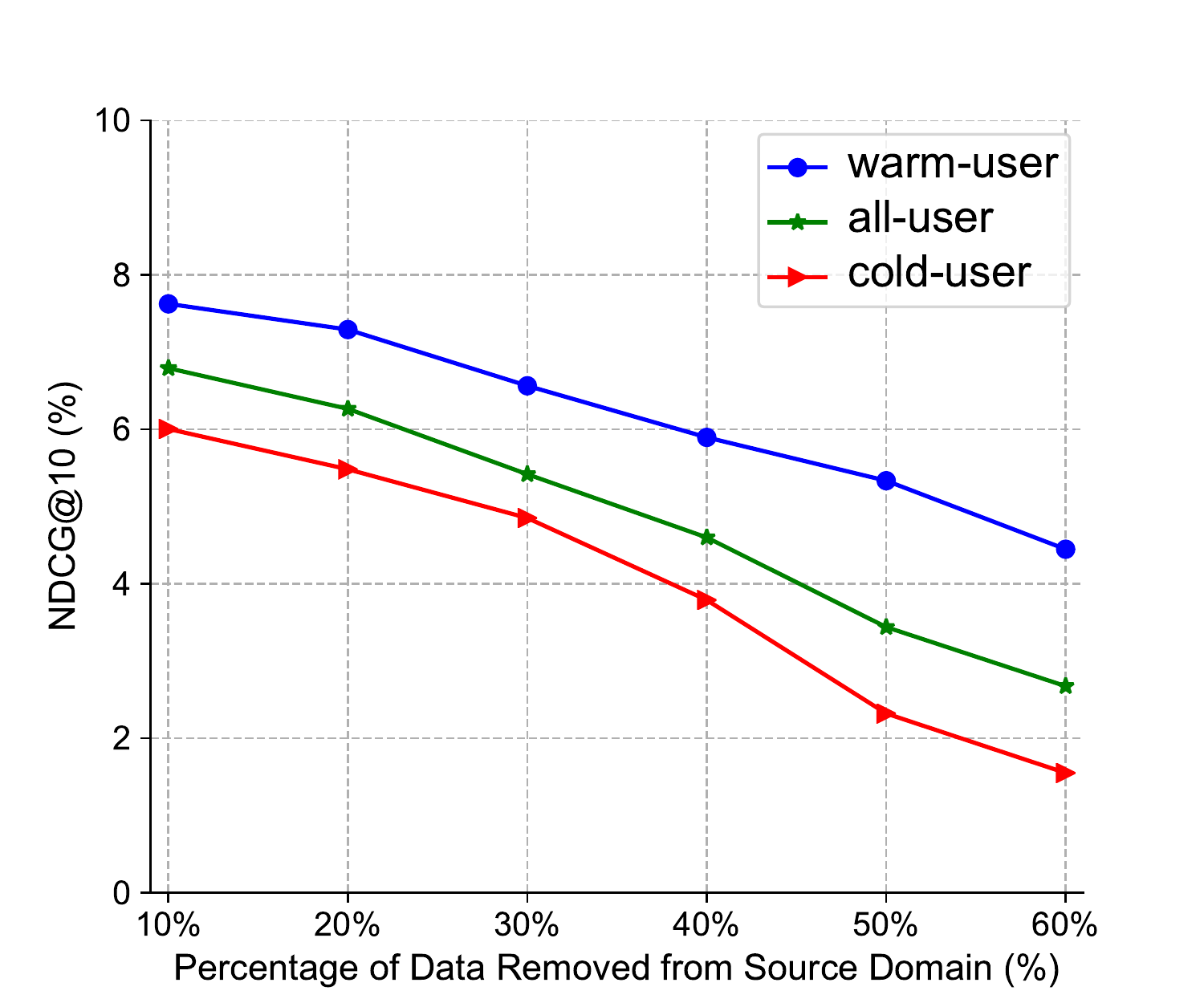}}
    \caption{The impact of the source domain data.}
    \label{fig:impact}
\end{figure}

Figure \ref{fig:impact} shows the impact of source domain data on the performance. We can see a clear drop on both HitRate and NDCG as the percentage of training data from the source domain decreases. Recommending cold items to cold users are more difficult as the overall performance is worse than recommending to users with interactions in target domain already. What is more, with less data from the source domain, the performance on cold-user recommendation decreases more than the warm-user situations with increasing gap between two lines. An interesting finding is that cold-user recommendation on Amazon data is more sensitive to the varying source data than that on TPMV data. We 
believe that it is related to the correlations between tags extracted from both domains. With less overlapping tags, it becomes more difficult to match interest between domains with progressively sparser source triggers.

\subsection{Visualization of User Group Preference}
\begin{figure}[h]
\setlength{\belowcaptionskip}{-0.15in}
    \centering
    \vspace{-0.1in}
    \subfigure[Clustering w/o orthogonality]{\includegraphics[scale=0.11]{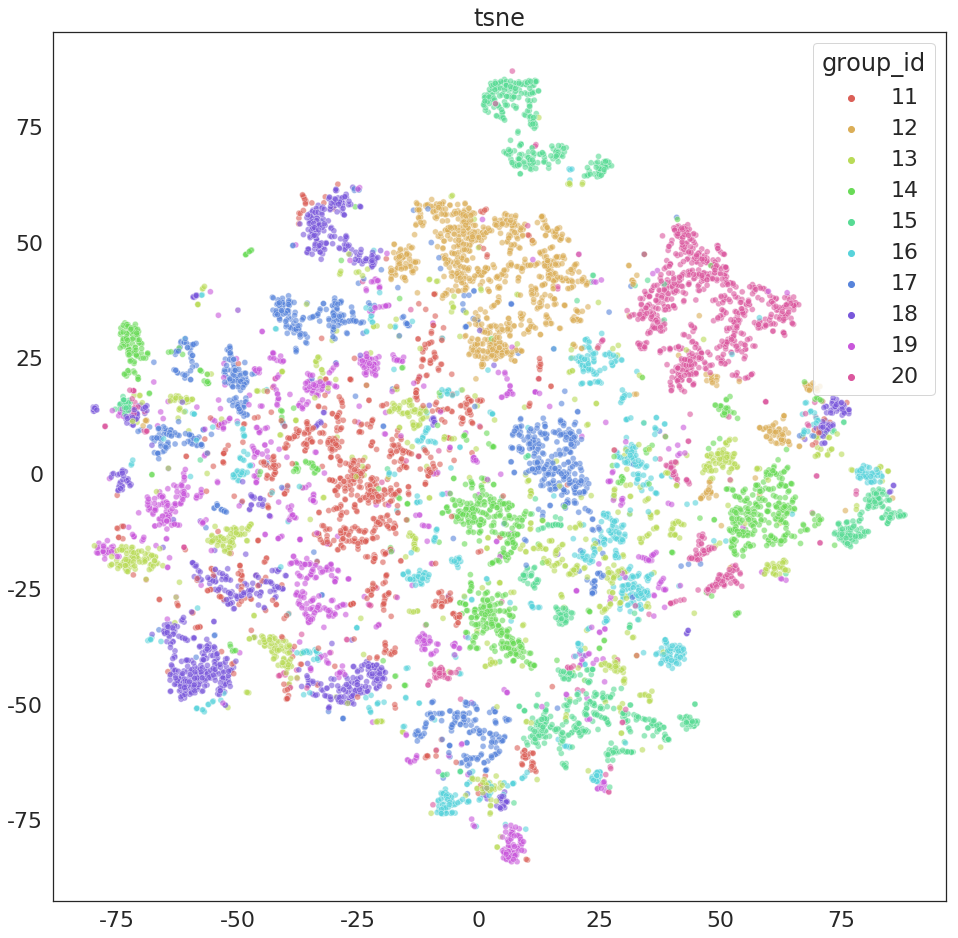}}
    \quad
    \vspace{-0.1in}
    \subfigure[Clustering w/ orthogonality]{\includegraphics[scale=0.11]{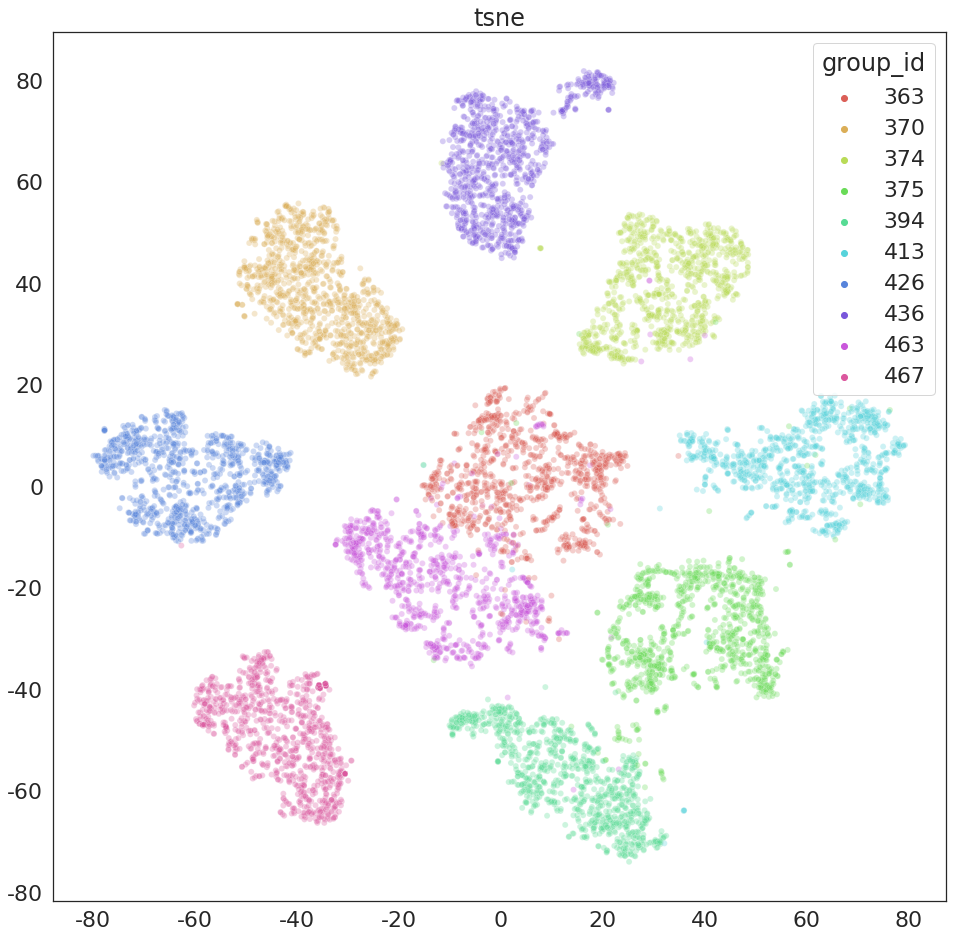}}
    \caption{t-SNE of clustering samples of the memory tree.}
    \label{fig:tsne}
\end{figure}

A good orthogonal representation tree should be able to project and classify users into distinct groups. To further verify whether orthogonal regularization works, We visualize the user group representations with t-distributed stochastic neighbor embedding (t-SNE)  \cite{van2008visualizing}. First, we obtain the memory slot vector of leaf nodes $\{\boldsymbol{s}\}_{k=1}^{h}$ via fully-connected projection of user history representation $\boldsymbol{e}_{u}^{(h)}$. Then the leaf node group $k$ with largest weight calculated using Eq. \eqref{nodeweight} is assigned to the user. Following t-SNE, We treat the each component of $\boldsymbol{s}_k$ as an individual point and keep only the two components that have the highest confidence levels. For better illustration, we randomly sample 10 user groups with 10,000 points within each and the result is shown in Figure \ref{fig:tsne}.

It is obvious that the clustering result with orthogonal regularization has higher similarity within a cluster and can separate different clusters better. When node vectors approach orthogonal, they become de-correlated so that the responses are much less redundant. By applying orthogonality constraints together with Gumbel-Softmax sampling trick, we observe a stable convergence and layer-wise distribution of the hierarchical memory tree.

\section{Online Deployment}
We conduct online experiments by deploying our work to guess-you-like recommendation after purchase at Taobao platform, where brand new micro-videos with extracted tags are displayed to users with related interest. This scenario has over dozens of millions of active users and hundreds of thousands of new micro-movies uploaded each day. For comparison, all the matching methods share the same ranking stage, so that click-through-rate (CTR) can be a fair metric and is used for evaluation. We compare our work with two baselines that are already running online, one is author-based CF where new videos created by the same author will be retrieved and the other is Youtube DNN where tag features are fed into the network for similarity measurement. Users assigned to the experiment group experience COUPLE recommendation, while the other two control groups experience author-CF and Youtube DNN, respectively. We run the A/B test in the heavy-traffic scenario for a week. The CTR of the experiment group with COUPLE improves over the control group by \textbf{8.855\%}. Further compared to author-CF, COUPLE has a lower CTR rate (-5.6\%) but provides significant gain on exposure rate by \textbf{58.91\%}. And It illustrates that with content-based recommendation scheme, more cold micro-videos get the opportunity to be shown to users which contributes to the sustainable development of recommender systems. In general COUPLE + author-CF improve CTR over Youtube + author-CF by \textbf{3.15\%}. We replace Youtube DNN with COUPLE and it is now serving online for the cross-domain cold micro-video recommendation.

\section{Conclusion}
In this paper, we proposed a unified framework for cross-domain cold-start recommendation in real world. It learns a user's preference from three different perspectives including user history, user content and user group. By jointly modeling the user's behaviors with extracted semantic tags across domains, our work is able to perform reliable recommendation in a new domain with limit user interactions.
Fed into a contrastive learning scheme, COUPLE achieve state-of-the-art results on two datasets and demonstrates superior performance compared to other baselines running at Taobao APP. It has been deployed in production at scale for cold micro-video recommendation.
For the future, we plan to integrate more heterogeneous content features into our framework to further help with the cold-start recommendation. Another exciting direction is to incorporate knowledge graph for information propagation towards explainable recommendation.

\balance
\bibliographystyle{ACM-Reference-Format}
\bibliography{ref}

\clearpage
\appendix
\section{Reproductivity Supplement}
In this section, we provide details on both offline and online experiment settings for reproductivity purpose. 

\subsection{Offline Experimental Details}
\subsubsection{Tag Extraction}
\begin{figure}[h]
\vspace{-0.05in}
    \centering
    \includegraphics[scale=0.36]{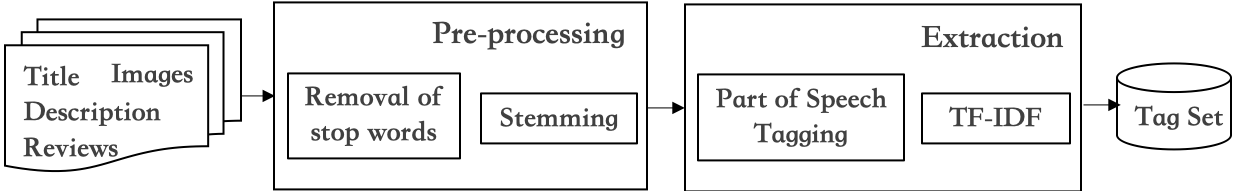}
    \caption{Tag extraction process used in paper.} \vspace{-0.15in}
    \label{fig:tag_process}
\end{figure}

The semantic tags used for representing items are extracted manually from different multi-media resources. For Amazon dataset, we extract meaningful short terms as tags from title, category, brand based on frequency and semantic annotation algorithms \cite{zhang2008comparative}. For TPMV dataset, richer tags can be drawn from product title, attributes, images via specialized in-house models. The tag extraction process can be referred to Figure \ref{fig:tag_process}.

\subsubsection{Parameter Settings of Our Method}
We implement our model with Tensorflow version 1.12 where default initialization recommended by Tensorflow is applied. We use dimension size 64 for all feature representations including tags, items and users. 

\textbf{User History Representation}. To model user history representation, one block of self-attention with hidden size 64 and heads of 4 is used. The tag numbers per item and item sequence length are capped to (10, 8) and (15, 5) for TPMV and Amazon datasets, respectively. Domain embedding is added to the item embedding to distinguish whether the items are from source or target domain.

\textbf{Memory Tree Implementation}. For the hierarchical memory tree design, we use 1-32-512 as the structure while dropout rate is set to be 0.2, the penalty parameter $\lambda$ for the SRIP regularization is set to be 0.1, and the temperature factor $\tau$ used in Gumbel-Softmax anneals using the schedule $\tau=\max(1,20{e}^{-rt})$ of the global training step $t$, where $\tau$ is updated every 1000 steps and $r=1\mathrm{e}-5$ are the hyperparameters used in our method.

\begin{figure}[b]
\twocolumn[{
    \centering
    \includegraphics[scale=0.49]{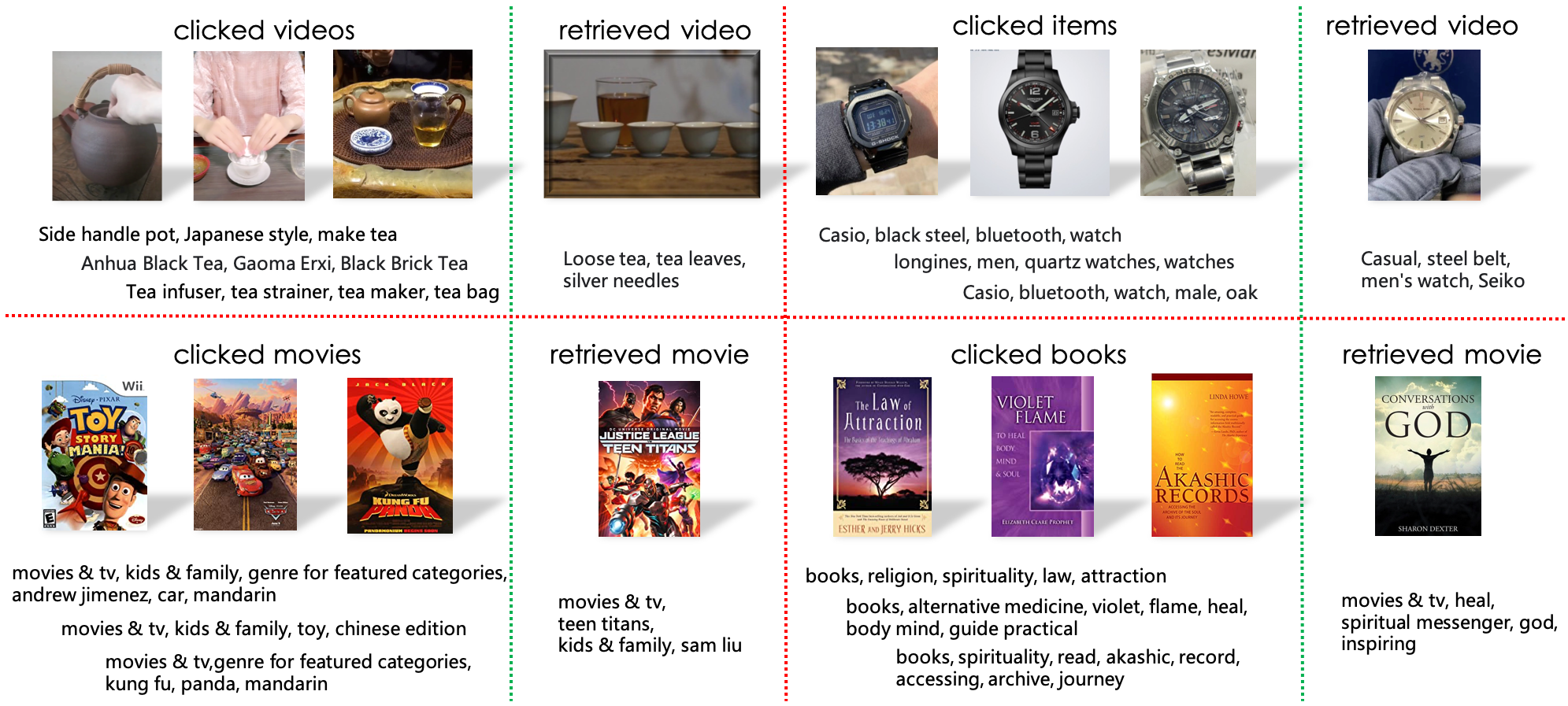}\vspace{-0.15in}
    \caption{Retrieval cases of our model.}
    \label{fig:cases}\vspace{0.15in}
}
]
\end{figure}

\textbf{Contrastive Learning}. The contrastive learning is performed with a FIFO queue of size 10$\times$256 where 256 is the mini-batch size. The temperature $\omega$ used in contrastive loss from Eq. \eqref{contrastive} is 0.03. The Adam \cite{kingma2014adam} optimizer for mini-batch gradient descent is used in our work and the learning rate is set to be 0.0001. Some extracted tags and retrieved cases of our work are visualized in Figure \ref{fig:cases}.

\subsection{Online Experimental Details}
\begin{figure}[h]
    \centering
    \includegraphics[scale=0.34]{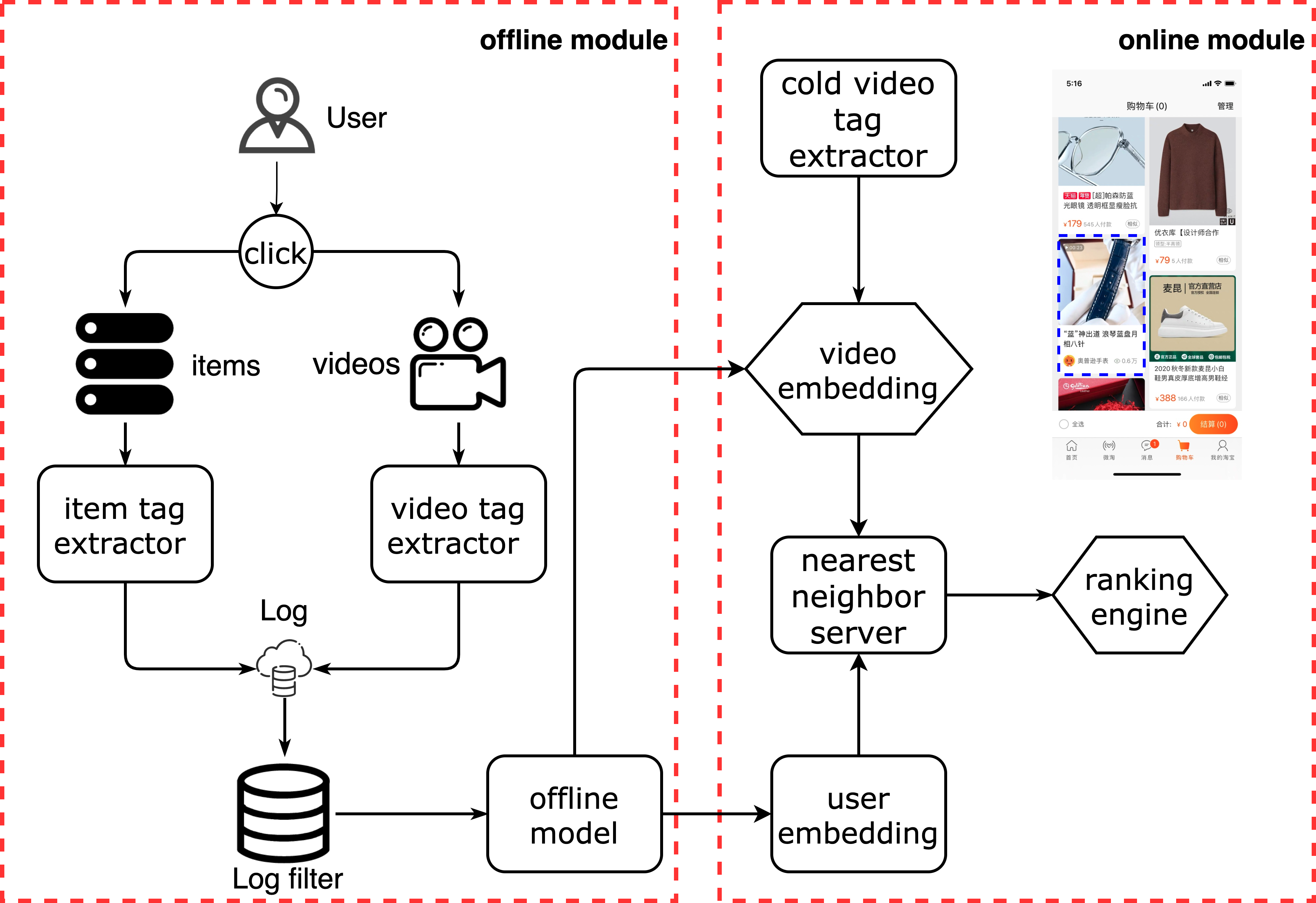}
    \caption{Online system workflow.}
    \label{fig:system}
\end{figure}

Our online system is shown in Figure \ref{fig:system}. The offline model trained with the user behaviors is utilized for online recommendation system. A user's representation is inferred with recent interacted items (represented by tags). On the other hand, new videos will go through a tag extractor and obtain the embeddings through the model. Real-time matching will be performed via calculating the similarity score between user and video embeddings (by inner product). Videos with the highest scores are retrieved for ranking stage. 

\end{document}